%% file: main.tex
\documentclass[preprint2]{aastex}
\usepackage{amsmath}
\usepackage{rotating}

\begin{document}
\title
{
  Toward tight gamma-ray burst luminosity relations
}
\author{Shi Qi$^{1,2,3}$ and Tan Lu$^{1,2,3}$}
\affil
{
  $^1$ Purple Mountain Observatory, Chinese Academy of Sciences,
  Nanjing 210008, China
}
\affil
{
  $^2$ Joint Center for Particle, Nuclear Physics and Cosmology,
  Nanjing University---Purple Mountain Observatory,
  Nanjing  210093, China
}
\affil
{
  $^3$ Key Laboratory of Dark Matter and Space Astronomy, Chinese
  Academy of Sciences, Nanjing 210008, China
}
\email{qishi11@gmail.com \quad t.lu@pmo.ac.cn}

\begin{abstract}
  The large scatters of luminosity relations of gamma-ray bursts
  (GRBs) have been one of the most important reasons that prevent the
  extensive applications of GRBs in cosmology.
  Many efforts have been made to seek tight luminosity relations.
  With the latest sample of $116$ GRBs with measured redshift and
  spectral parameters, we investigate $6$
  two-dimensional (2D) correlations and $14$ derived
  three-dimensional (3D) correlations of GRBs to explore the
  possibility of decreasing the intrinsic scatters of the luminosity
  relations of GRBs.
  We find the 3D correlation of
  $E_{\mathrm{peak}}$--$\tau_{\mathrm{RT}}$--$L$
  to be evidently tighter
  (at the $2 \sigma$ confidence level)
  than its corresponding 2D correlations,
  i.e., the $E_{\mathrm{peak}}$--$L$ and $\tau_{\mathrm{RT}}$--$L$
  correlations.
  In addition, the coefficients before the logarithms of
  $E_{\mathrm{peak}}$ and $\tau_{\mathrm{RT}}$
  in the $E_{\mathrm{peak}}$--$\tau_{\mathrm{RT}}$--$L$ correlation
  are almost exact opposites of each other.
  Inputting this situation as a prior reduces the relation to
  $L \propto
  (E_{\mathrm{peak}}^{'} / \tau_{\mathrm{RT}}^{'})^{0.842 \pm 0.064}$,
  where $E_{\mathrm{peak}}^{'}$ and $\tau_{\mathrm{RT}}^{'}$ denote the
  peak energy and minimum rise time in the GRB rest frame.
  We discuss how our findings can be interpreted/understood in the
  framework of the definition of the luminosity (energy released in
  units of time).
  Our argument about the connection between the luminosity relations
  of GRBs and the definition of the luminosity provides a clear
  direction for exploring tighter luminosity relations of GRBs in the
  future.
\end{abstract}


\section{Introduction}

Gamma-ray bursts (GRBs) are the most luminous astrophysical events
observed so far. The high luminosities make them observable at very
high redshifts. The measured highest redshift for GRBs has exceeded
$8$ (e.g., the GRB 090423 with a redshift of
$z \approx 8.2$~\citep{Tanvir:2009ex, Salvaterra:2009ey}
and the GRB 090429B with a photometric redshift of
$z \sim 9.4$~\citep{Cucchiara:2011pj}).
Many efforts have been made to apply them to cosmology~\citep[see,
for example,][etc.]{Dai:2004tq, Ghirlanda:2004fs, Firmani:2005gs,
  Lamb:2005cw, Liang:2005xb, Xu:2005uv, Wang:2005ic, Ghirlanda:2006ax,
  Schaefer:2006pa, Wang:2007rz, Li:2007re, Amati:2008hq,
  Basilakos:2008tp, Qi:2008zk, Qi:2008ag, Kodama:2008dq, Liang:2008kx,
  Wang:2008vja, Qi:2009yr, Cardone:2009mr, Izzo:2009bw, Liang:2009gw,
  Cardone:2010rr, Wang:2011bx}.
The key for this is to find some luminosity relation that relates
the luminosity (e.g., the isotropic peak luminosity $L$) or energy
(e.g., the isotropic energy $E_{\gamma, \mathrm{iso}}$ or the
collimation-corrected energy $E_{\gamma}$) of GRBs to their measurable
properties (the luminosity indicators).
The tighter the relation is, the more accurate and reliable
information about the universe we can get from it.
Many luminosity relations have been discovered in the literature,
e.g., the relations of
$\tau_{\mathrm{lag}}$--$L$~\citep{Norris:1999ni},
$V$--$L$~\citep{Fenimore:2000vs, Reichart:2000kq},
$E_{\mathrm{peak}}$--$E_{\gamma, \mathrm{iso}}$~\citep{Amati:2002ny},
$E_{\mathrm{peak}}$--$E_{\gamma}$~\citep{Ghirlanda:2004me},
$E_{\mathrm{peak}}$--$L$~\citep{Schaefer:2002tf, Wei:2002gb,
  Yonetoku:2003gi},
and
$\tau_{\mathrm{RT}}$--$L$~\citep{Schaefer:2006pa}, etc.
Here, the spectral lag $\tau_{\mathrm{lag}}$ is the time shift between
the hard and soft light curves.
The variability $V$ is a quantitative measurement of the spikiness of
the light curve, which can be obtained by calculating the normalized
variance of the observed light curve around the smoothed light curve.
There exist several definitions of $V$, depending mainly on the
smoothing time intervals upon which the reference curve is built and
the normalization.
$E_{\mathrm{peak}}$ is the photon energy at which the $\nu F_{\nu}$
spectrum peaks.
The minimum rise time $\tau_{\mathrm{RT}}$ of a GRB light curve is the
shortest time over which the light curve rises by half the peak flux
of the pulse.
In addition, correlations were also found between the transition times
of the X-ray light curve from exponential to power law and the X-ray
luminosities at the transitions~\citep{Dainotti:2008vw,
  Dainotti:2010ki, Qi:2009zs}.

The most crucial problem with GRBs for cosmological application is
that their luminosity relations are usually quite scattered, which
prevents them from being used as good standard candles as Type Ia
supernovae.
Therefore, one of the most important directions of cosmological study
for GRBs is to seek GRB luminosity relations with
sufficiently small intrinsic scatters.
Many attempts have been made in literature,
among which, a remarkable approach is to explore possible hidden
parameters in known correlations in order to reduce the intrinsic
scatters.
For example, \citet{Firmani:2006gw} claimed that a temporal parameter
of the prompt emission, $T_{0.45}$, could reduce the scatter of
the correlation of $E_{\mathrm{peak}}$--$L$ to a negligible value.
It was later found that the new proposed relation does not appear
to be as tight as it seemed~\citep{Rossi:2008mv, Collazzi:2008gu}.
\citet{Tsutsui:2008sy, Tsutsui:2010mg} suggested that introducing the
parameter
$T_L \equiv E_{\gamma, \mathrm{iso}} / L$
to the $E_{\mathrm{peak}}$--$L$ relation
could substantially reduce the intrinsic scatters of the correlation
of $E_{\mathrm{peak}}$--$L$ and
$E_{\mathrm{peak}}$--$E_{\gamma, \mathrm{iso}}$.
\citet{Xu:2011bv} reported that a significantly tighter correlation
can be obtained by adding $E_{\gamma, \mathrm{iso}}$ to the correlation
discovered in~\citet{Dainotti:2008vw, Dainotti:2010ki}.
However, they did not perform the normalization (see the following
section of this paper for details about normalization) before
comparing the intrinsic scatters. Taking into account the
normalization, the new relation is in fact looser instead of tighter
in the sense of cosmological distance measurement.
In~\citet{Yu:2009xd}, more general analysis was carried out,
investigating $5$ two-dimensional (2D) correlations and $10$ derived
three-dimensional (3D) correlations to explore the
possibility of reducing the intrinsic scatters.

Since we have more GRBs at hand now~\citep{Xiao:2009dr, Wang:2011bx},
it is beneficial to update the comparison between 2D and 3D
luminosity relations of GRBs and to check whether the conclusions
drawn from an earlier GRB sample still hold.
This is the main aim of this paper.
We included more GRB luminosity relations in our analysis than there
are in~\citet{Yu:2009xd}.
Six 2D correlations and $14$ derived 3D correlations were
investigated.
We basically followed the method used in~\citet{Yu:2009xd} for the
analysis.
To be complete, we briefly describe the method in the following
section.
The results and discussion are presented in Section 3.
A summary is given in the last section.

\section{Methodology}

For the 2D luminosity relations, we consider the following
correlations:
\begin{align}
  \label{eq:GRB-lag-L}
  \log \frac{L}{1 \; \mathrm{erg} \; \mathrm{s}^{-1}}
  &= a_1+b_1 \log
  \left[
    \frac{\tau_{\mathrm{lag}}(1+z)^{-1}}{0.1\;\mathrm{s}}
  \right]
  ,
  \\
  \label{eq:GRB-V-L}
  \log \frac{L}{1 \; \mathrm{erg} \; \mathrm{s}^{-1}}
  &= a_2+b_2 \log
  \left[
    \frac{V(1+z)}{0.02}
  \right]
  ,
  \\
  \label{eq:GRB-E_peak-L}
  \log \frac{L}{1 \; \mathrm{erg} \; \mathrm{s}^{-1}}
  &= a_3+b_3 \log
  \left[
    \frac{E_{\mathrm{peak}}(1+z)}{300\;\mathrm{keV}}
  \right]
  ,
  \\
  \label{eq:GRB-E_peak-E_gamma}
  \log \frac{E_{\gamma}}{1\;\mathrm{erg}}
  &= a_4+b_4 \log
  \left[
    \frac{E_{\mathrm{peak}}(1+z)}{300\;\mathrm{keV}}
  \right]
  ,
  \\
  \label{eq:GRB-tau_RT-L}
  \log \frac{L}{1 \; \mathrm{erg} \; \mathrm{s}^{-1}}
  &= a_5+b_5 \log
  \left[
    \frac{\tau_{\mathrm{RT}}(1+z)^{-1}}{0.1\;\mathrm{s}}
  \right]
  ,
  \\
  \label{eq:GRB-E_peak-E_iso}
  \log \frac{E_{\gamma, \mathrm{iso}}}{1\;\mathrm{erg}}
  &= a_6+b_6 \log
  \left[
    \frac{E_{\mathrm{peak}}(1+z)}{300\;\mathrm{keV}}
  \right]
  .
\end{align}
The first five of the above correlations are exactly the
luminosity relations considered in~\citet{Yu:2009xd}.
With the latest compilation of GRBs, the $V$--$L$ relation has become
quite scattered~\citep{Xiao:2009dr, Wang:2011bx}, which makes the
parameter $V$ a very poor luminosity indicator. Despite this, we still
include this relation in our analysis in order to investigate whether
$V$ is helpful in reducing the intrinsic scatter when constructing 3D
luminosity relations together with other luminosity indicators.
In this paper, we added in our analysis the correlation of
Eq.~(\ref{eq:GRB-E_peak-E_iso}), for which more GRBs can be utilized,
compared with the similar correlation of
Eq.~(\ref{eq:GRB-E_peak-E_gamma}), due to its independence from the
GRB jet opening angle.
In the correlations, the isotropic peak luminosity $L$, the isotropic
energy $E_{\gamma, \mathrm{iso}}$, and the collimation-corrected
energy $E_{\gamma}$ are derived from the observables of the bolometric
peak flux $P_{\mathrm{bolo}}$, the bolometric fluence
$S_{\mathrm{bolo}}$, and the beaming factor $F_{\mathrm{beam}}$ through
\begin{align}
  \label{eq:GRB-L-P_bolo}
  L &= 4 \pi d_L^2 P_{\mathrm{bolo}}
  ,
  \\
  \label{eq:GRB-E_iso-S_bolo}
  E_{\gamma,\mathrm{iso}}
  &=
  4 \pi d_L^2 S_{\mathrm{bolo}} (1+z)^{-1}
  ,
  \\
  \label{eq:GRB-E_gamma-S_bolo}
  E_{\gamma}
  &=
  E_{\gamma,\mathrm{iso}} F_{\mathrm{beam}}
  ,
\end{align}
where $d_L$ is the luminosity distance, which depends on the
cosmological model and is inversely proportional to the value of
the Hubble parameter of today.
As in~\citet{Yu:2009xd}, we adopt the flat $\Lambda$CDM model with
$\Omega_m=0.27$ and replace $d_L$ with
$\bar{d}_L = \frac{H_0}{c} d_L \times 1 \, \mathrm{cm}$
in the calculation, so that the dependence on the Hubble constant is
absorbed into the intercepts $a_i$ of the linear luminosity
relations.

The method used to extend the 2D luminosity relations to 3D ones is
also basically the same as that adopted in~\citet{Yu:2009xd}, except
we have more 3D luminosity relations here due to the addition of
the correlation of Eq.~(\ref{eq:GRB-E_peak-E_iso}).
First, for later convenience when denoting specific correlations,
we summarize the equations for the 2D luminosity relations above as
\begin{equation}
  \label{eq:GRB-2D}
  y^{(i)} = c_0^{(i, \, i)} + c_1^{(i, \, i)} x^{(i)}
  ,
\end{equation}
where
\begin{equation}
  x^{(1)}
  =
  \log
  \left[
    \frac{\tau_{\mathrm{lag}}(1+z)^{-1}}{0.1\;\mathrm{s}}
  \right]
  ,
\end{equation}
\begin{equation}
  x^{(2)}
  =
  \log
  \left[
    \frac{V(1+z)}{0.02}
  \right]
  ,
\end{equation}
\begin{equation}
  x^{(3)}
  =
  x^{(4)} = x^{(6)} =
  \log
  \left[
    \frac{E_{\mathrm{peak}}(1+z)}{300\;\mathrm{keV}}
  \right]
  ,
\end{equation}
\begin{equation}
  x^{(5)}
  =
  \log
  \left[
    \frac{\tau_{\mathrm{RT}}(1+z)^{-1}}{0.1\;\mathrm{s}}
  \right]
  ,
\end{equation}
\begin{equation}
  y^{(1)}
  =
  y^{(2)} = y^{(3)} = y^{(5)} =
  \log \frac{L}{1 \; \mathrm{erg} \; \mathrm{s}^{-1}}
  ,
\end{equation}
\begin{equation}
  y^{(4)}
  =
  \log \frac{E_{\gamma}}{1\;\mathrm{erg}}
  ,
\end{equation}
\begin{equation}
  y^{(6)}
  =
  \log \frac{E_{\gamma, \mathrm{iso}}}{1\;\mathrm{erg}}
  ,
\end{equation}
and
\begin{equation}
  c_0^{(i, \, i)} = a_i, \quad c_1^{(i, \, i)} = b_i
  .
\end{equation}
The coefficients $c$ are given two superscripts to incorporate the 3D
correlations introduced below.
We denote a luminosity relation by the superscript pair of the
corresponding $c$.
We let $i < j$ for all the 3D luminosity relations $(i, \, j)$ to
avoid duplication, and we classify them into the following three
groups for convenience:
\begin{enumerate}
\item \emph{Correlations between the luminosity (the isotropic peak
    luminosity $L$) and two luminosity indicators.}
  For $(i, \, j)$ with both $i$ and $j$ in $(1, \, 2, \, 3, \, 5)$,
  the 3D luminosity relations are
  \begin{equation}
    \label{eq:GRB-3D_1}
    y^{(i)} = c_0^{(i, \, j)}
    + c_1^{(i, \, j)} x^{(i)}
    + c_2^{(i, \, j)} x^{(j)}
    .
  \end{equation}
\item \emph{Correlations between energy (the isotropic energy
    $E_{\gamma, \mathrm{iso}}$ or the collimation-corrected energy
    $E_{\gamma}$) and two luminosity indicators.}
  For $(i, \, j) = (1, \, 4)$, $(2, \, 4)$, or $(4, \, 5)$, the 3D
  luminosity relations are
  \begin{equation}
    \label{eq:GRB-3D_2.1}
    y^{(4)} = c_0^{(i, \, j)}
    + c_1^{(i, \, j)} x^{(i)}
    + c_2^{(i, \, j)} x^{(j)}
    ,
  \end{equation}
  and for $(i, \, j) = (1, \, 6)$, $(2, \, 6)$, or $(5, \, 6)$ they
  are
  \begin{equation}
    \label{eq:GRB-3D_2.2}
    y^{(6)} = c_0^{(i, \, j)}
    + c_1^{(i, \, j)} x^{(i)}
    + c_2^{(i, \, j)} x^{(j)}
    .
  \end{equation}
\item \emph{Correlations between the luminosity, the energy
    $E_{\gamma, \mathrm{iso}}$ or $E_{\gamma}$, and the peak energy
    $E_{\mathrm{peak}}$.}
  For $(i, \, j) = (3, \, 4)$ or $(3, \, 6)$, the 3D luminosity
  relations are
  \begin{equation}
    \label{eq:GRB-3D_3}
    y^{(i)} = c_0^{(i, \, j)}
    + c_1^{(i, \, j)} x^{(i)}
    + c_2^{(i, \, j)} y^{(j)}
    .
  \end{equation}
\end{enumerate}

A luminosity relation can be multiplied by an equal constant on both
sides of the equation without actually changing the correlation
itself.
However, the multiplication would change the intrinsic scatter of the
correlation by a factor of the absolute value of the constant
multiplied on the equation.
To compare the intrinsic scatters of the luminosity relations, we
divide both sides of Eq.~(\ref{eq:GRB-3D_3}) by a factor of
$1 - c_2^{(i, \, j)}$, so that $\log (d_L)$ terms have the same
coefficient in all the luminosity relations discussed here.
We normalize the luminosity relations in this way because,
in addition to being helpful in understanding the GRBs themselves,
the luminosity relations are mainly aimed at distance measurements.

To explore possible hidden parameters in the 2D luminosity relations,
we compare the intrinsic scatters of 3D luminosity relations with
those of corresponding 2D ones. The principle is that the two
luminosity relations compared with each other should have two
parameters in common, so that we can conclude whether the intrinsic
scatter is reduced by introducing the third parameter.
Following this principle, for the 3D
luminosity relations $(i, \, j)$ in the first and third classes,
i.e., those of Eq.~(\ref{eq:GRB-3D_1}) and~(\ref{eq:GRB-3D_3}),
they are compared with 2D luminosity relations $(i, \, i)$ and
$(j, \, j)$, and for the 3D luminosity relations in the second class,
i.e., those of Eq.~(\ref{eq:GRB-3D_2.1}) and~(\ref{eq:GRB-3D_2.2}),
they are compared with 2D luminosity
relations $(4, \, 4)$ and $(6, \, 6)$ respectively.
Only when the intrinsic scatter of a 3D luminosity relation is smaller
than that of \emph{all} its corresponding 2D luminosity relation(s)
can we say that the intrinsic scatter is reduced.

In the fit of the luminosity relations, we used the techniques
presented in~\citet{Agostini:2005fe}.
Using the correlation of Eq.~(\ref{eq:GRB-3D_1}) as an example, the
joint likelihood function for the coefficients $c$ and the intrinsic
scatter $\sigma_{\mathrm{int}}$ is given as
\begin{align}
  \label{eq:likelihood}
  \mathcal{L}(c, \, \sigma_{\mathrm{int}})
  &\propto \prod_k
  \frac{1}
  {
    \sqrt
    {
      \sigma_{\mathrm{int}}^2
      + \sigma_{y_k^{(i)}}^2
      + c_1^2 \sigma_{x_k^{(i)}}^2
      + c_2^2 \sigma_{x_k^{(j)}}^2
    }
  }
  \nonumber \\
  &
  \times
  \exp
  \left[
    -
    \frac
    {
      \left(
        y_k^{(i)}
        - c_0
        - c_1 x_k^{(i)}
        - c_2 x_k^{(j)}
      \right)^2
    }
    {
      2
      \left(
        \sigma_{\mathrm{int}}^2
        + \sigma_{y_k^{(i)}}^2
        + c_1^2 \sigma_{x_k^{(i)}}^2
        + c_2^2 \sigma_{x_k^{(j)}}^2
      \right)
    }
  \right]
  ,
\end{align}
where $k$ runs over GRBs with corresponding quantities available.
The joint likelihood functions are similar for other 3D luminosity
relations. For the 2D luminosity relations, the joint likelihood
functions can be obtained by setting $c_2 = 0$.

For the GRB data, we used the compilation in~\citet{Wang:2011bx},
which includes $116$ GRBs. For this compilation, the data of the
luminosity indicators are taken from~\citet{Xiao:2009dr}.
The size of this GRB sample is almost twice as large as the
compilation in~\citet{Schaefer:2006pa}, which was used in the previous
study to compare 2D and 3D luminosity relations~\citep{Yu:2009xd}.
In addition, significant improvements have been made in the
calculation of the luminosity indicators for the updated compilation.
When considering error propagation from a quantity, say $\xi$ with
error $\sigma_{\xi}$, to its logarithm, we set
$
[
  \log(\xi + \sigma_{\xi}^{+})
  +
  \log(\xi - \sigma_{\xi}^{-})
]
/ 2
$
and
$
[
  \log(\xi + \sigma_{\xi}^{+})
  -
  \log(\xi - \sigma_{\xi}^{-})
]
/ 2
$
as the center value and the error of the logarithm correspondingly.
This requires $\xi > \sigma_{\xi}^{-}$ (the quantities we are
interested in here are all positive).
Due to the limitation of the data,
for a given luminosity relation $(i, \, j)$, not all the GRBs have all
of the needed observational quantities available and satisfy
$\xi > \sigma_{\xi}^{-}$ at the same time. By set $(i, \, j)$
we denote the maximum GRB set that can be used in the luminosity
relation $(i, \, j)$.
The numbers of GRBs of different sets are presented in
Table~\ref{tab:fit}.

\section{Results and discussion}

We summarize our results for the fits to the luminosity relations and
comparisons between 2D and 3D correlations in Table~\ref{tab:fit}.
We list our findings from the table as follows:
\begin{enumerate}
\item For the correlations common with those studied
  in~\citet{Yu:2009xd}, including both 2D and 3D ones, almost all the
  intrinsic scatters have increased with the updated GRB
  sample. For the 2D ones, only the $\tau_{\mathrm{RT}}$--$L$ relation
  retains a comparable intrinsic scatter.
\item For the cases of $(1, \, 3)$ and $(1, \, 5)$, the intrinsic
  scatters of the 3D correlations are reduced at the $1 \sigma$
  confidence level compared with their corresponding 2D correlations,
  and for the case of $(3, \, 5)$, the intrinsic scatter is reduced at
  the $2 \sigma$ confidence level. No other statistically significant
  reduction of the intrinsic scatters is found.
\item For the case of $(3, \, 5)$, the values of the coefficients
  $c_1$ and $c_2$ are almost exact opposites of each other. For the
  case of $(1, \, 3)$, the values of $c_1$ and $c_2$ are also rough
  opposites of each other.
\end{enumerate}

\begin{sidewaystable*}[htbp]
  \centering
  \tiny
  \noindent
  \makebox[\textwidth][c]
  {
    \begin{tabular}{l @{} c @{} c @{} c @{} c @{} c @{} c}
      \hline \hline
      \\
      $(i, j)$ & 1 & 2 & 3 & 4 & 5 & 6 \\
      \\
      & $\tau_{\mathrm{lag}}$ & $V$ & $E_{\mathrm{peak}}$
      & $E_{\mathrm{peak}}$ & $\tau_{\mathrm{RT}}$
      & $E_{\mathrm{peak}}$ \\
      \hline
      \\
      \input{table}
    \end{tabular}
  }
  \caption
  {
    Fit of 2D and 3D luminosity relations.
    The first row and the first column of the table are the indices
    denoting the luminosity relations and the corresponding luminosity
    indicators.
    In every cell of the table, the first row is the luminosity and/or
    energy involved in the relations which, together with the
    luminosity indicators, tell what the correlations are about.
    The second row of every cell is the number of GRBs of
    set $(i, \, j)$,
    the vector below enclosed by parentheses is the vector of
    $c$ for the luminosity relation $(i, \, j)$, and what follows
    next is the intrinsic scatter.
    For 3D luminosity relations, the reduction of the intrinsic
    scatters compared with corresponding 2D luminosity relations is
    presented in the brackets.
    The statistics in the table are for the median values and the
    errors with the $1 \sigma$ ($68.3\%$) confidence level.
  }
  \label{tab:fit}
\end{sidewaystable*}

Item 1 reminds us once again of the complexity of GRBs. Correlations
should be tested against a large enough sample to determine how good
they are. The intrinsic scatter derived from a small sample may be
significantly affected by the selection effect of the sample itself.
Item 2 is basically consistent with the previous studies
of~\citet{Yu:2009xd}, in which the intrinsic scatters are reduced at
$1 \sigma$ and $2 \sigma$ confidence levels, respectively, for the
cases of $(1, \, 3)$ and $(3, \, 5)$, i.e.,
the $\tau_{\mathrm{lag}}$--$E_{\mathrm{peak}}$--$L$
and $E_{\mathrm{peak}}$--$\tau_{\mathrm{RT}}$--$L$ correlations.
With the new GRB sample adopted here,
the intrinsic scatter is also reduced at the $1 \sigma$ confidence
level for the case of $(1, \, 5)$, i.e., the
$\tau_{\mathrm{lag}}$--$\tau_{\mathrm{RT}}$--$L$ correlation,
the edges of the two $1 \sigma$
confidence intervals of the magnitude of reduction of the intrinsic
scatter for $(1, \, 5)$ are both very close to zero, which reduces its
reliability.
The correlation $(3, \, 5)$ turns
out to be the only 3D correlation that appears very robust in the test
of the reduction of the intrinsic scatter.

The correlation $(3, \, 5)$ is among $E_{\mathrm{peak}}$,
$\tau_{\mathrm{RT}}$, and $L$. It is a relation among an energy scale,
a timescale, and the luminosity.
This naturally makes us think of the definition of the luminosity
(energy released in units of time).
As can be seen from the large scatters of the 2D luminosity relations,
it is most likely that, due to their complexity, the luminosities of
GRBs cannot be well determined from only one quantity.
However, according to the definition of the luminosity,
we should in principle be able to calculate the
luminosity from a characteristic energy scale and a characteristic
timescale of GRBs.
If we could find two measurable quantities (e.g., an energy scale and
a timescale) that are strongly correlated with the assumed
characteristic energy scale and timescale, then we would
find a 3D luminosity relation between the luminosity and the two
measurable quantities.
The quality of the relation obviously depends on the strength of the
correlations between the two quantities and the assumed characteristic
energy scale and timescale, and is also related to the degree of match
between the two quantities in the sense of the match between energy
scale and timescale in the definition of the luminosity
(i.e., the energy scale and timescale chosen to calculate the
luminosity should correspond to each other).
The intrinsic scatter of such a 3D luminosity relation is expected to
be significantly reduced compared with its corresponding 2D ones.
The situation described in item 2 suggests $E_{\mathrm{peak}}$ and
$\tau_{\mathrm{RT}}$ as one such pair of measurable quantities.

\begin{figure}[tbp]
  \centering
  \includegraphics[width = 0.5 \textwidth]{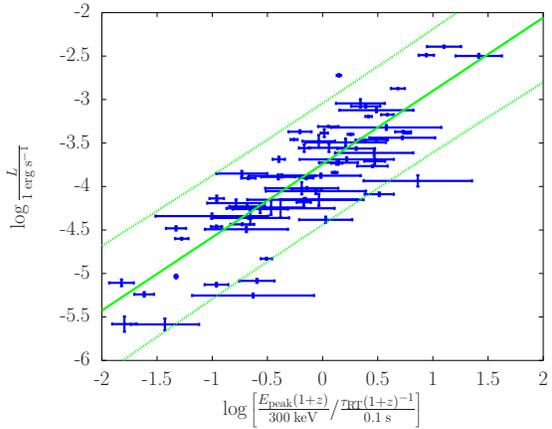}
  \caption
  {
    Best fit of the GRBs to the luminosity relation of
    Eq.~(\ref{eq:35_2d}) and the corresponding $2 \sigma$ confidence
    region.
    In the calculation of the $L$, the dependence of the luminosity
    distance on the Hubble constant is absorbed into the intercept of
    the relation.
  }
  \label{fig:fig35_2d}
\end{figure}

What makes the correlation $(3, \, 5)$ more interesting is the
situation described in item 3.
Following the discussion above, the almost exact opposition of $c_1$
and $c_2$ suggests a considerably high degree of match between
$E_{\mathrm{peak}}$ and $\tau_{\mathrm{RT}}$ considering that, in the
definition of the luminosity, the indices for the energy scale and the
timescale are opposite.
In comparison, the degree of match between $E_{\mathrm{peak}}$ and
$\tau_{\mathrm{lag}}$ is slightly weaker.
The situation of the almost exact opposite for the correlation
$(3, \, 5)$ does not show up in the results of~\citet{Yu:2009xd}.
However, remember that, for the GRB data used here, not only is the
sample size much larger, but significant improvements also have
been made in the calculation of the luminosity indicators, so we have
reasons to treat the situation as more than just a coincidence.
If we input the opposition of $c_1$ and $c_2$ as a prior for the
correlation, then the relation is reduced to a 2D one, that is,
\begin{equation}
  \label{eq:35_2d}
  \log \frac{L}{1 \; \mathrm{erg} \; \mathrm{s}^{-1}}
  = a + b \log
  \left[
    \frac{E_{\mathrm{peak}}(1+z)}{300\;\mathrm{keV}}
    /
    \frac{\tau_{\mathrm{RT}}(1+z)^{-1}}{0.1\;\mathrm{s}}
  \right]
  .
\end{equation}
A fit to this relation gives the result of
$a = -3.742_{-0.048}^{+0.047}$,
$b = 0.842_{-0.064}^{+0.064}$,
and the intrinsic scatter
$\sigma_{\mathrm{int}} = 0.346_{-0.032}^{+0.037}$.
We present the relation in Figure~\ref{fig:fig35_2d}.
Using $E_{\mathrm{peak}}^{'}$ and $\tau_{\mathrm{RT}}^{'}$ to denote the
peak energy and minimum rise time in the GRB rest frame, the relation
can be expressed as
\begin{equation}
  \label{eq:35_2d_rest}
  L \propto
  \left(
    \frac{E_{\mathrm{peak}}^{'}}{\tau_{\mathrm{RT}}^{'}}
  \right)^{0.842 \pm 0.064}
  .
\end{equation}
Following the definition of the luminosity, if appropriate luminosity
is used on the left-hand side of Eq.~(\ref{eq:35_2d}) or
Eq.~(\ref{eq:35_2d_rest}), the coefficient $b$ in Eq.~(\ref{eq:35_2d})
or the index in Eq.~(\ref{eq:35_2d_rest}) should be $1$.
However, in practice, there are various definitions of luminosities,
e.g., peak or averaged luminosity and isotropic or
collimation-corrected luminosity.
It is unclear which luminosity is attached to the pair of energy scale
and timescale used on the right-hand side of equations,
i.e., $E_{\mathrm{peak}}$ and $\tau_{\mathrm{RT}}$.
What we adopt in the relations is the isotropic peak luminosity $L$.
Thus, the coefficient $b$ in Eq.~(\ref{eq:35_2d}) or the index in
Eq.~(\ref{eq:35_2d_rest}) may be interpreted as the correction factor
between the luminosity attached to the pair of $E_{\mathrm{peak}}$ and
$\tau_{\mathrm{RT}}$ and the isotropic peak luminosity.

For a clear and complete physical interpretation of the
$E_{\mathrm{peak}}$--$\tau_{\mathrm{RT}}$--$L$ correlation, it should
be explained how $E_{\mathrm{peak}}$ and $\tau_{\mathrm{RT}}$ are
correlated with the assumed energy scale and timescale.
This is an issue that needs further detailed study, which is out of
the scope of this short paper.
Here, we present some general speculation on this.
One may note that $E_{\mathrm{peak}}$ is not a radiated energy, but
the photon energy at which the $\nu F_{\nu}$ spectrum peaks.
The calculation of the luminosity needs the radiated energy.
$E_{\mathrm{peak}}$ may relate itself with a radiated energy through a
physical process that releases photons with the frequency at which
$\nu F_{\nu}$ spectrum peaks, or it may just be a dominant parameter
in the spectrum that can be used as a representation of the bolometric
energy released by the GRB, as can be seen from
$E_{\mathrm{peak}}$--$E_{\gamma, \mathrm{iso}}$ and
$E_{\mathrm{peak}}$--$E_{\gamma}$ correlations
($E_{\gamma, \mathrm{iso}}$ and $E_{\gamma}$ cannot replace the role
of $E_{\mathrm{peak}}$ here, mostly because their calculation
depends on the luminosity distance and the intrinsic scatters are
enlarged in the procedure of normalization, while $E_{\mathrm{peak}}$
is directly measured from the observation and is independent of the
cosmological model).
We think that $\tau_{\mathrm{RT}}$ stands out in our investigation
most likely because $\tau_{\mathrm{RT}}$ is the
\emph{minimum} rise time.
Though requiring it to be minimum sounds like a very simple operation
in the data processing, it could be a very important step for
extracting the clean physics from the complex GRBs, thus eliminating
intrinsic scatters from many irrelevant physical processes.

We would also like to mention the 3D
correlation of $(3, \, 6)$, i.e.,
the $E_{\mathrm{peak}}$--$E_{\gamma, \mathrm{iso}}$--$L$ correlation
and the $E_p$--$T_L$--$L_p$ correlation
referenced in~\citet{Tsutsui:2008sy, Tsutsui:2010mg}.
Using the notation adopted here, the latter is given by
\begin{equation}
  \label{eq:GRB-E_peak-T_L-L}
  \log L = A + B \log [ E_{\mathrm{peak}} (1+z) ] + C \log T_L
  ,
\end{equation}
where $T_L = E_{\gamma,\mathrm{iso}} / L$.
Substituting $T_L$ into the equation, the correlation can be rewritten
as
\begin{equation}
  \label{eq:GRB-E_peak-T_L-L-1}
  (1+C) \log L = A + B \log [ E_{\mathrm{peak}} (1+z) ]
  + C \log E_{\gamma,\mathrm{iso}}
  .
\end{equation}
Comparing it with Eq.~(\ref{eq:GRB-3D_3}), we can see that this
relation is in fact the same as the 3D luminosity relation
$(3, \, 6)$, and it has
been already normalized.
However, the conclusion we draw here for the correlation conflicts
with that in~\citet{Tsutsui:2008sy, Tsutsui:2010mg}.
\citet{Tsutsui:2008sy, Tsutsui:2010mg} claimed that the intrinsic
scatter of the correlation $E_p$--$T_L$--$L_p$ is smaller than that of
both the $E_{\mathrm{peak}}$--$L$ and
$E_{\mathrm{peak}}$--$E_{\gamma, \mathrm{iso}}$ correlations.
Our result shows that the intrinsic scatter of the correlation of
$(3, \, 6)$ is not reduced compared with its corresponding 2D
correlations. In fact, its intrinsic scatter even appears a little
larger.
The advantages of our analysis here are that we used more GRBs,
which helped us reduce the selection effect from the sample itself,
and that we adopted a sophisticated statistical method,
in which not only the intrinsic scatter but also its error can be
derived, so that we can judge the statistical significance of the
reduction of the intrinsic scatters.

\section{Summary}

As a further step on in seeking tight luminosity relations of
GRBs, we fitted the latest data of GRBs with measured redshift and
spectral parameters to $6$ 2D correlations and $14$
derived 3D correlations and compared their intrinsic scatters to
explore possible hidden parameters in the 2D correlations.
Compared with the analysis of an early sample of GRBs, the intrinsic
scatters of most of the luminosity relations have increased,
which reminds us of the complexity of GRBs.
Correlations should be tested against a large enough sample to reduce
the possible selection effects from the sample itself.
Our result shows that the
$E_{\mathrm{peak}}$--$\tau_{\mathrm{RT}}$--$L$
correlation appears to be significantly tighter
(at the $2 \sigma$ confidence level)
than its corresponding 2D correlations,
i.e., the $E_{\mathrm{peak}}$--$L$ and $\tau_{\mathrm{RT}}$--$L$
correlations.
What is more interesting is that,
in the $E_{\mathrm{peak}}$--$\tau_{\mathrm{RT}}$--$L$ correlation,
the coefficients before the logarithms of $E_{\mathrm{peak}}$ and
$\tau_{\mathrm{RT}}$ are almost exact opposites of each other.
If this situation is input as a prior, the correlation is reduced to
a 2D one, $L \propto
(E_{\mathrm{peak}}^{'} / \tau_{\mathrm{RT}}^{'})^{0.842 \pm 0.064}$,
where $E_{\mathrm{peak}}^{'}$ and $\tau_{\mathrm{RT}}^{'}$ denote the
peak energy and minimum rise time in the GRB rest frame.
We interpret/understand the result in the framework of the definition
of the luminosity (energy released in units of time).
Due to the complexity of GRBs, it is unlikely that we can accurately
determine their luminosity through only one quantity, as can be
seen from the large scatters of the 2D luminosity relations.
And in principle, we should be able to calculate the luminosity from a
characteristic energy scale and a characteristic timescale of GRBs.
Our result suggests that $E_{\mathrm{peak}}$ and $\tau_{\mathrm{RT}}$
are a pair of measurable quantities which are sufficiently correlated
with the assumed characteristic energy scale and timescale and that
there is a high degree of match between them (in the sense of the
match between energy scale and timescale in the definition of the
luminosity),
so we may construct a better luminosity relation with them.
Since there are different definitions of the luminosity, the index of
$0.842$ in
$L \propto
(E_{\mathrm{peak}}^{'} / \tau_{\mathrm{RT}}^{'})^{0.842 \pm 0.064}$
may be interpreted as the correction factor between the luminosity
attached to the pair of $E_{\mathrm{peak}}$ and $\tau_{\mathrm{RT}}$
and the isotropic peak luminosity $L$.

Our argument about the connection between the luminosity relations
of GRBs and the definition of the luminosity provides a clear
direction for exploring tighter luminosity relations of GRBs in the
future.
It should also be easier to seek correlations between measurable
quantities and the assumed characteristic energy scale and timescale
related to the luminosity than to directly seek the relations between
the measurable quantities and the luminosity.
Our findings about the $E_{\mathrm{peak}}$--$\tau_{\mathrm{RT}}$--$L$
correlation illustrate this approach.

\acknowledgments

We thank the anonymous referee for helpful comments and suggestions.
This work was supported by the National Natural Science Foundation of
China under grant no.~10973039, the Chinese Academy of Sciences under
grant no.~KJCX2-EW-W01, and the China Postdoctoral
Science Foundation under grant no.~20100471421 (for Shi Qi).

\end{document}

%% file: table.tex
& $L$ & $L$ & $L$ & $E_{\gamma}$ & $L$ & $E_{\gamma, \mathrm{iso}}$ \\
\\
1 & $53$ & $46$ & $53$ & $13$ & $49$ & $46$ \\
\\
$\tau_{\mathrm{lag}}$ &
$\left( -3.958_{-0.067}^{+0.067}, \, -0.77_{-0.10}^{+0.10} \right)$ &
$\left( -4.04_{-0.13}^{+0.13}, \, -0.70_{-0.14}^{+0.14}, \, 0.14_{-0.15}^{+0.15} \right)$ &
$\left( -4.014_{-0.059}^{+0.059}, \, -0.57_{-0.10}^{+0.10}, \, 0.67_{-0.16}^{+0.17} \right)$ &
$\left( -5.565_{-0.094}^{+0.093}, \, -0.06_{-0.16}^{+0.16}, \, 1.33_{-0.25}^{+0.24} \right)$ &
$\left( -3.750_{-0.079}^{+0.081}, \, -0.41_{-0.12}^{+0.12}, \, -0.71_{-0.17}^{+0.17} \right)$ &
$\left( -3.471_{-0.074}^{+0.074}, \, -0.43_{-0.12}^{+0.12}, \, 0.88_{-0.19}^{+0.19} \right)$ \\
\\
&
$0.476_{-0.046}^{+0.054}$ &
$0.479_{-0.050}^{+0.060}$ &
$0.402_{-0.041}^{+0.048}$ &
$0.25_{-0.08}^{+0.10}$ &
$0.404_{-0.041}^{+0.049}$ &
$0.452_{-0.049}^{+0.059}$ \\
\\
&
&
$\left[ -0.002_{-0.075}^{+0.073}, \, 0.192_{-0.080}^{+0.078} \right]$ &
$\left[ 0.074_{-0.066}^{+0.068}, \, 0.135_{-0.061}^{+0.059} \right]$ &
$\left[ 0.05_{-0.12}^{+0.11} \right]$ &
$\left[ 0.072_{-0.067}^{+0.068}, \, 0.069_{-0.064}^{+0.064} \right]$ &
$\left[ 0.037_{-0.069}^{+0.065} \right]$ \\
\hline
\\

& & $L$ & $L$ & $E_{\gamma}$ & $L$ & $E_{\gamma, \mathrm{iso}}$ \\
\\
2 & & $81$ & $81$ & $21$ & $64$ & $68$ \\
\\
$V$ &
\nodata &
$\left( -4.39_{-0.14}^{+0.14}, \, 0.60_{-0.15}^{+0.15} \right)$ &
$\left( -4.31_{-0.11}^{+0.11}, \, 0.33_{-0.13}^{+0.13}, \, 1.13_{-0.15}^{+0.15} \right)$ &
$\left( -5.71_{-0.14}^{+0.14}, \, 0.14_{-0.14}^{+0.14}, \, 1.45_{-0.16}^{+0.16} \right)$ &
$\left( -3.68_{-0.14}^{+0.14}, \, 0.20_{-0.13}^{+0.13}, \, -1.03_{-0.15}^{+0.15} \right)$ &
$\left( -3.62_{-0.13}^{+0.13}, \, 0.26_{-0.13}^{+0.13}, \, 1.10_{-0.16}^{+0.16} \right)$ \\
\\
&
&
$0.672_{-0.053}^{+0.060}$ &
$0.503_{-0.041}^{+0.046}$ &
$0.207_{-0.055}^{+0.067}$ &
$0.428_{-0.040}^{+0.046}$ &
$0.483_{-0.042}^{+0.049}$ \\
\\
&
&
&
$\left[ 0.168_{-0.070}^{+0.073}, \, 0.034_{-0.060}^{+0.059} \right]$ &
$\left[ 0.096_{-0.096}^{+0.099} \right]$ &
$\left[ 0.244_{-0.070}^{+0.072}, \, 0.045_{-0.062}^{+0.062} \right]$ &
$\left[ 0.007_{-0.061}^{+0.060} \right]$ \\
\hline
\\

& & & $L$ & $L$, $E_{\gamma}$ & $L$ & $L$, $E_{\gamma, \mathrm{iso}}$ \\
\\
3 & & & $116$ & $24$ & $72$ & $101$ \\
\\
$E_{\mathrm{peak}}$ &
\nodata &
\nodata &
$\left( -4.134_{-0.053}^{+0.053}, \, 1.40_{-0.12}^{+0.12} \right)$ &
$\left( -1.3_{-1.3}^{+1.3}, \, 0.49_{-0.38}^{+0.38}, \, 0.46_{-0.24}^{+0.23} \right)$ &
$\left( -3.739_{-0.062}^{+0.061}, \, 0.84_{-0.12}^{+0.12}, \, -0.85_{-0.11}^{+0.11} \right)$ &
$\left( -1.24_{-0.27}^{+0.27}, \, 0.15_{-0.14}^{+0.14}, \, 0.827_{-0.075}^{+0.074} \right)$ \\
\\
&
&
&
$0.538_{-0.038}^{+0.042}$ &
$0.78_{-0.25}^{+0.60}$ &
$0.348_{-0.033}^{+0.038}$ &
$2.2_{-0.7}^{+1.6}$ \\
\\
&
&
&
&
$\left[ -0.24_{-0.60}^{+0.25}, \, -0.48_{-0.60}^{+0.26} \right]$ &
$\left[ 0.189_{-0.053}^{+0.054}, \, 0.125_{-0.056}^{+0.058} \right]$ &
$\left[ -1.7_{-1.6}^{+0.7}, \, -1.8_{-1.6}^{+0.7} \right]$ \\
\hline
\\

& & & & $E_{\gamma}$ & $E_{\gamma}$ & \\
\\
4 & & & & $24$ & $22$ & \\
\\
$E_{\mathrm{peak}}$ &
\nodata &
\nodata &
\nodata &
$\left( -5.639_{-0.074}^{+0.071}, \, 1.47_{-0.20}^{+0.19} \right)$ &
$\left( -5.60_{-0.11}^{+0.11}, \, 1.40_{-0.25}^{+0.24}, \, -0.12_{-0.23}^{+0.22} \right)$ &
\nodata \\
\\
&
&
&
&
$0.304_{-0.068}^{+0.082}$ &
$0.342_{-0.074}^{+0.092}$ &
\\
\\
&
&
&
&
&
$\left[ -0.04_{-0.11}^{+0.11} \right]$ &
\\
\hline
\\

& & & & & $L$ & $E_{\gamma, \mathrm{iso}}$ \\
\\
5 & & & & & $72$ & $62$ \\
\\
$\tau_{\mathrm{RT}}$ &
\nodata &
\nodata &
\nodata &
\nodata &
$\left( -3.563_{-0.074}^{+0.074}, \, -1.12_{-0.14}^{+0.14} \right)$ &
$\left( -3.316_{-0.082}^{+0.081}, \, -0.32_{-0.15}^{+0.15}, \, 0.99_{-0.17}^{+0.16} \right)$ \\
\\
&
&
&
&
&
$0.473_{-0.042}^{+0.048}$ &
$0.469_{-0.043}^{+0.050}$ \\
\\
&
&
&
&
&
&
$\left[ 0.020_{-0.063}^{+0.061} \right]$ \\
\hline
\\

& & & & & & $E_{\gamma, \mathrm{iso}}$ \\
\\
6 & & & & & & $101$ \\
\\
$E_{\mathrm{peak}}$ &
\nodata &
\nodata &
\nodata &
\nodata &
\nodata &
$\left( -3.532_{-0.053}^{+0.052}, \, 1.47_{-0.12}^{+0.12} \right)$ \\
\\
&
&
&
&
&
&
$0.490_{-0.038}^{+0.042}$ \\
\hline